\documentclass[sigconf]{acmart}

\renewcommand\footnotetextcopyrightpermission[1]{} 
\usepackage{booktabs} 
\usepackage{cleveref}
\usepackage{listings}
\usepackage{xcolor}
\usepackage{colortbl}

\definecolor{verylightgray}{rgb}{.97,.97,.97}
\definecolor{lightgray}{rgb}{.95,.95,.95}

\setcopyright{rightsretained}

\acmDOI{10.475/123_4}

\acmISBN{123-4567-24-567/08/06}

\acmConference[WETSEB]{International Workshop on Emerging Trends in Software Engineering for Blockchain}{2018}{Gothenburg, Sweden}
\acmYear{2018}
\copyrightyear{2018}

\acmArticle{4}
\acmPrice{15.00}

\begin{document}
\title{\textsc{ReviewChain}: Untampered Product Reviews on the Blockchain}

\author{Daniel Martens}
\affiliation{%
  \institution{University of Hamburg}
  \city{Hamburg}
  \state{Germany}
}
\email{martens@informatik.uni-hamburg.de}

\author{Walid Maalej}
\affiliation{%
  \institution{University of Hamburg}
  \city{Hamburg}
  \state{Germany}
}
\email{maalej@informatik.uni-hamburg.de}

\renewcommand{\shortauthors}{Martens et al.}

\begin{abstract}
Online portals include an increasing amount of user feedback in form of ratings and reviews. 
Recent research highlighted the importance of this feedback and confirmed that positive feedback improves product sales figures and thus its success. 
However, online portals' operators act as central authorities throughout the overall review process. 
In the worst case, operators can exclude users from submitting reviews, modify existing reviews, and introduce fake reviews by fictional consumers.
This paper presents \textsc{ReviewChain}, a decentralized review approach.
Our approach avoids central authorities by using blockchain technologies, decentralized apps and storage.
Thereby, we enable users to submit and retrieve untampered reviews.
We highlight the implementation challenges encountered when realizing our approach on the public Ethereum blockchain.
For each implementation challange, we discuss possible design alternatives and their trade-offs regarding costs, security, and trustworthiness.
Finally, we analyze which design decision should be chosen to support specific trade-offs and present resulting combinations of decentralized blockchain technologies, also with conventional centralized technologies.
\end{abstract}

%
%



\maketitle

\section{Introduction}

Online portals often allow consumers to rate products on a five-star scale and by writing a review message.
Ratings and reviews are used for various purposes, e.g., for products bought on Amazon\footnote{https://www.amazon.com}, trips booked on Tripadvisor\footnote{https://www.tripadvisor.com}, or apps downloaded from the Apple App Store\footnote{https://itunes.apple.com/us/genre/ios/id36?mt=8}.
These portals receive an increasing amount of user feedback, e.g., Tripadvsior claims to serve more than 500 million traveler-generated reviews.
Consumers use this feature to express their satisfaction and experiences with products.
A large body of research analyzes reviews, for example to summarize them~\cite{hu2004mining} and to understand their helpfulness~\cite{mudambi2010research}. 
When deciding between products, reviews have a considerable impact on consumers' choice~\cite{chatterjee2001online, gretzel2008use}.
Research found that ratings and review correlate with sales and download ranks~\cite{pagano2013user, harman2012app}.
Stable and numerous ratings are associated with high quality and lead to higher downloads and sales numbers.

Consumers that rely on reviews when deciding for a product, have to trust at least two parties involved.
These are the review authors and the operators of online portals.
Untrustworthy reviews of single authors, e.g., an extremely positive review in between negative reviews, can possibly be recognized by consumers.
However, this does not apply for larger amounts of reviews modified by online portal operators themselves.
The operators act as central authorities throughout the complete review process. 
In worst case, an operator can exclude consumers from submitting reviews, modify existing reviews, and introduce fake reviews to improve the ratings and rankings of products~\cite{fornaciari2014identifying,jindal2008opinion, mukherjee2013yelp}.

In this paper, we aim to resolve the problem of central authorities, being able to influence the review processes, by using the public Ethereum block\-chain.
Consumers no longer need to rely on central authorities as block\-chains operate decentrally across a network of several nodes, in which every user can participate~\cite{akcora2017blockchain}.
Thereby, we enable consumers to submit and access untampered and trustworthy reviews.
The \textbf{contribution} of the paper is twofold:

\begin{enumerate}
\item We propose a decentralized review approach which does not involve central authorities.
\item We summarize the approach's major implementation challenges on the public Ethereum blockchain, discuss design alternatives, and compare their trade-offs.
\end{enumerate}

The remainder of the papers is structured as follows:
Section~2 introduces the most important terminologies used in this paper.
Section 3 describes our \textsc{ReviewChain} approach and its design goals.
Section 4 discusses the major challenges encountered when implementing the approach on the Ethereum blockchain.
Finally, Section~5 analyzes configurations of design alternatives with regard to their resulting trade-offs.


\section{Terminology}

\noindent\textbf{Ethereum}\footnote{https://www.ethereum.org} is a blockchain, developed in 2014 by Vitalik Buterin.
In comparison to the Bitcoin blockchain, which only handles accounts and transactions, Ethereum also stores programming logic.
When paying for its execution, any turing-complete script can be run on Ethereum.
Thereby, it enables decentralized apps without any possibility of downtime, censorship, or third-party interference.

\noindent\textbf{Transactions} are digitally signed messages that are persisted on the blockchain. 
Each transaction is associated with an action, such as transferring cryptocurrency units.
Transactions have a sender and receiver address.
The addresses represent a public key belonging to a specific user.

\noindent\textbf{Ether} (ETH) is the cryptocurrency traded on the Ethereum block\-chain.
It is used to pay nodes of the network for executing requested operations, such as transactions.

\noindent\textbf{Smart Contracts} were first introduced by Szabo~\cite{szabo1994smart} in 1994.
These are self-executing and operate autonomously.
In Ethereum, smart contracts can be written by, e.g., using the Solidity programming language~\cite{wood2014ethereum}.
Contracts are executed on several nodes of the network within Ethereum virtual machines (EVM).
After executing a contract, nodes must reach a consensus of the calculated result.


\section{ReviewChain Approach}

\begin{figure}[]
\includegraphics[width=\columnwidth]{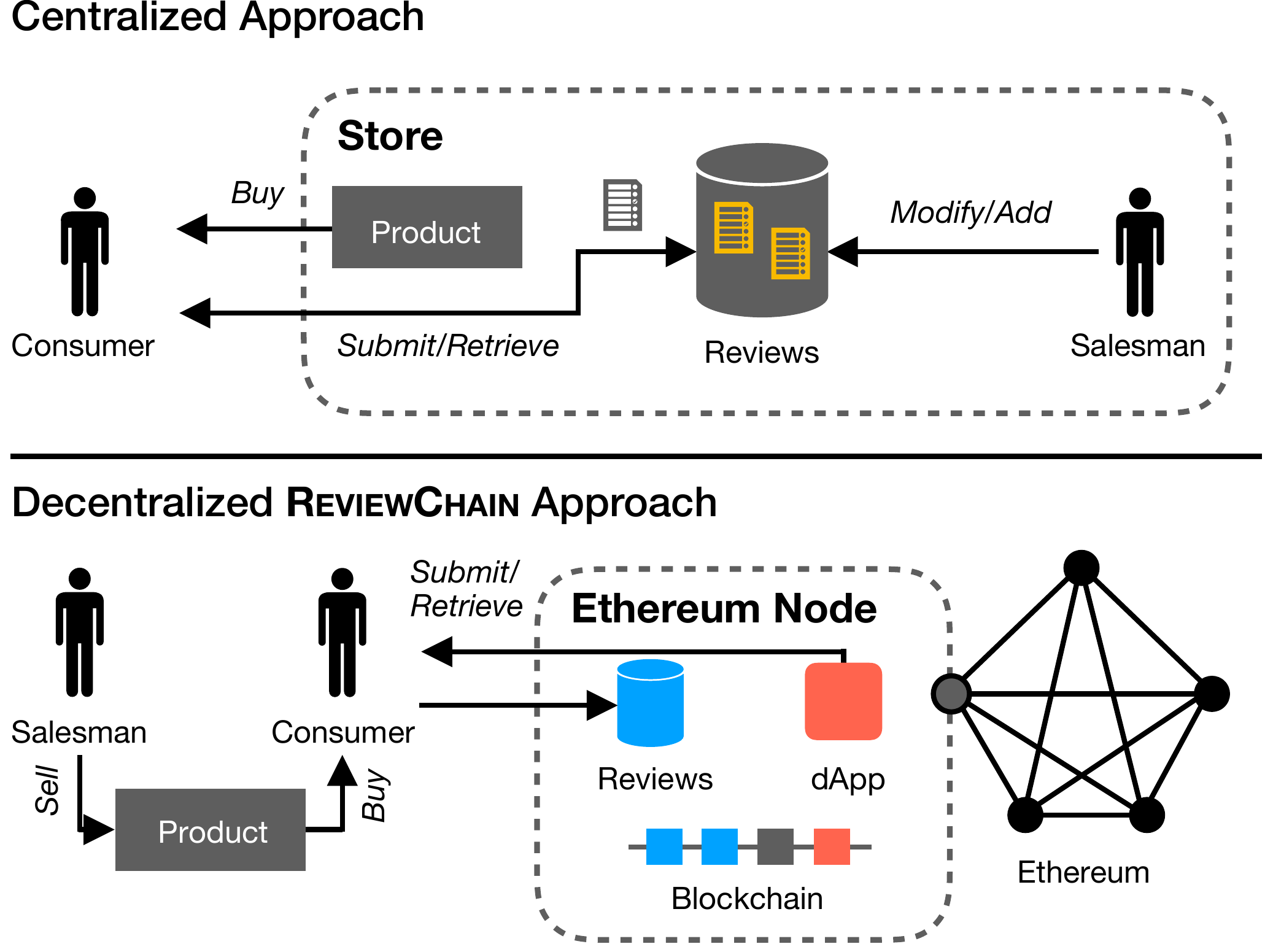}
\caption{Centralized and \textsc{ReviewChain} Approach.}
\label{fig:approach}
\end{figure}

\textsc{ReviewChain} focuses on creating and providing persistent access to untampered product ratings and reviews.
Its fundamental goal is to avoid central authorities.
Therefore, our approach utilizes technologies, such as the blockchain as decentralized and immutable data storage, and decentralized apps to provide access to reviews.

A comparison of the conventional centralized review approach and our decrentralized \textsc{ReviewChain} approach is shown in \Cref{fig:approach}.
In the upper part of the figure, central authorities are able to exclude specific consumers from submitting reviews.
Further, authorities store reviews in a centralized database.
This enables them to modify exisiting reviews and introduce fake reviews by fictional consumers that did not buy the reviewed product.
In the decrentralized approach, shown in the bottom part of the figure, this is no longer possible. Once stored on the blockchain, reviews cannot be modified, neither can consumers be excluded from providing reviews.
We refer to these reviews as \textit{untampered reviews}.

Our \textsc{ReviewChain} approach must fulfill four requirements. 
These are extracted from existing review functionalities used for product reviews on Amazon or app reviews within the Apple App Store:

\begin{enumerate}
\item Authors must have purchased the reviewed product,
\item Review authors must be distinguishable,
\item Each author can only submit one review per product version,
\item Review submissions must not result in any costs for authors.
\end{enumerate}

In the following, we apply our approach to the real-world scenario of mobile app reviews.
Therefore, we enable authors to submit reviews within native apps, and provide access to reviews via a web service.
By doing so, we aim to highlight additional challanges which result from the usage of different technological environments.
Moreover, we implement our approach on Ethereum to demonstrate the applicability and limitations of a blockchain technologies.


\section{Implementation Challenges}

This section describes the major implementation challenges of our approach. 
The challenges are divided into sections.
Each section describes possible design alternatives and discusses their trade-offs.


\subsection{Submitting Transactions via Mobile Apps}

\begin{figure}[b]
\includegraphics[width=\columnwidth]{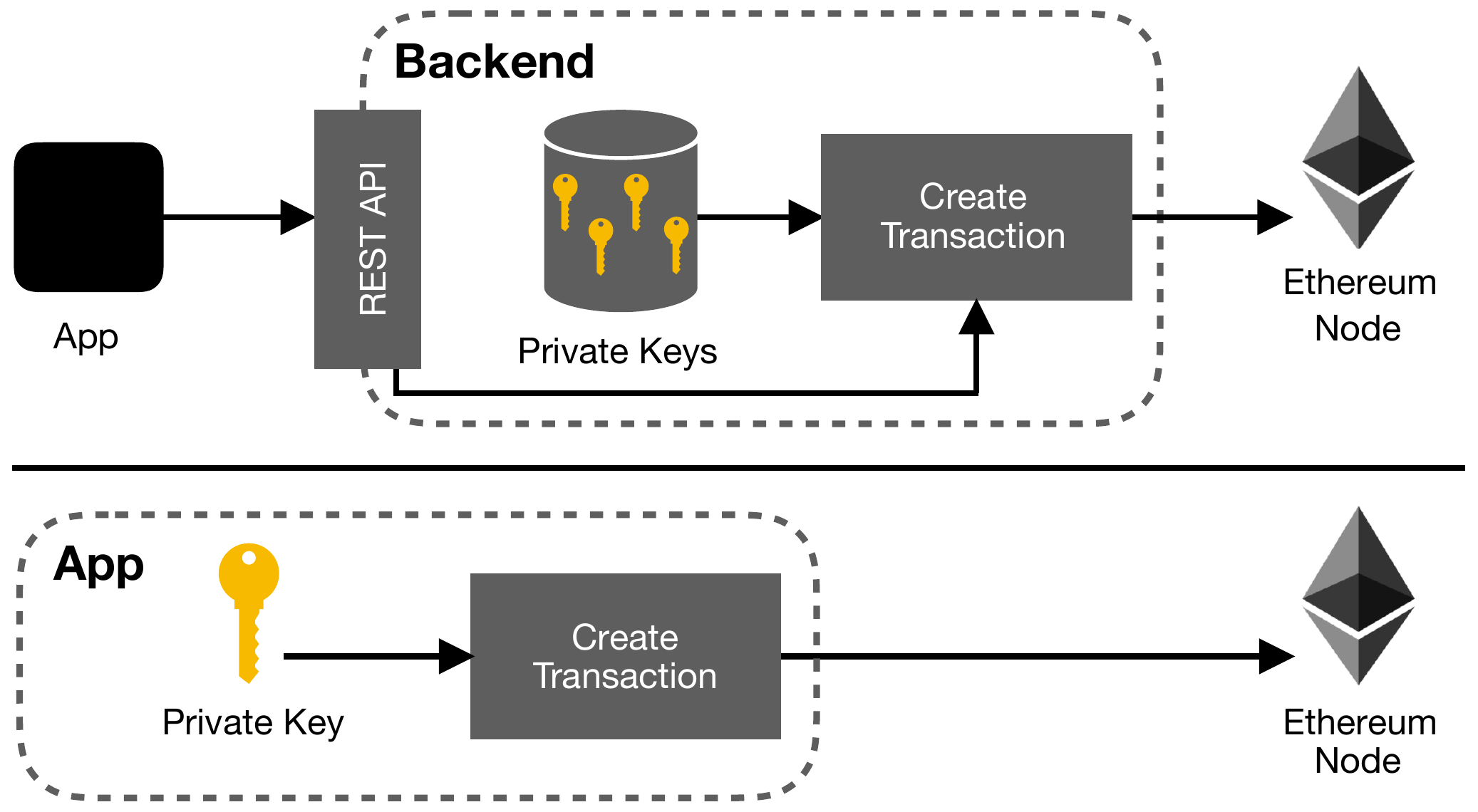}
\caption{Blockchain Transactions via Mobile Apps.}
\label{fig:execute}
\end{figure}

After publishing a smart contract on the blockchain, which allows consumers to store reviews, the contract is executed by sending a transaction to the contract address.
Each transaction needs to be signed with the user's private key.
\Cref{fig:execute} shows two alternatives to submit transactions via mobile apps.

The first alternative implements the transaction handling within a backend.
App and backend communicate using a REST API.
The transactions, as well as the private keys, are generated by the backend.
The keys are stored within a centralized database.
After signing, the transactions are sent to an Ethereum node.
This node can be run by the backend itself.
In the second alternative, the creation of transactions and private keys are contained within the app itself.
Signed transactions are directly sent to an Ethereum node.

Comparing both, the first alternative offers several advantages.
Computation-intensive operations, such as generating and de-/en\-crypting private keys can be performed on backend-side instead of using a resource-limited mobile device.
Also, as fees have to be paid to execute transactions and our approach requires reviewers not to pay for these, the central database of the backend can be used to create a list of all participating users' addresses.
These can be funded beforehand with tokens to pay the transaction fees of review submissions.
However, this alternative introduces a major drawback, the backend acts as a central authority.
By not authenticating users, operators can exclude those from submitting reviews.
Also, the stored private keys can be used to submit reviews without users' behalf.
As a result, a decentralized design, as presented in the second alternative, is preferable. 
In this design, the mobile app must self-contain the private key, as well as the transaction handling.

To realize the second alternative, an official implementation of the Ethereum protocol, called Go Ethereum (Geth)\footnote{https://geth.ethereum.org}, exists.
The implementation can be cross-compiled to both Android and iOS.
Using Java and Swift wrappers, it can be directly called from within the app's native implementation.
In addition, Geth generates native wrappers for smart contracts.
Geth offers account management, remote node interfacing, and enables interactions with smart contracts.
For example, private keys are generated and stored in an encrypted keystore.
This keystore offers a standard and light security mode.
However, due to the mobile device's resource limitations, the author suggest to use only the light security mode \cite{szilagyi2017accountmanagement}.
Although decentralized, this introduces a security risk.
The keys are stored in the app's document directory.
On regular devices, both Android and iOS, this folder can only be accessed by the application itself.
On jailbroken devices apps can access all files on the mobile device.
Thereby, the user's private keys can be transmitted to an external server.
A server's high computational capacities can be used to brute-force the private key's passphrase and perform transactions afterwards, without the behalf of the user.


\subsection{Restricting Access to App Users}

After a user submits a review, the transaction including the review is created, signed, and sent to an Ethereum node.
When the nodes of the network reached consensus, the transaction is immutably persisted on the blockchain.
Since each person can participate in the network, by downloading a copy of the blockchain and running a node, transactions, as well as smart contracts, are publicly visible.
This enables participants of the network to copy, modify (e.g., by changing the review text), and resubmit transactions.
The participants must not necessarily have downloaded the app the review targets. 
Thereby, fake reviews can be introduced to the blockchain.

Unfortunately, it is difficult to solve this issue without using a central authority.
In the following, we describe three alternatives to exclude non-app users from submitting reviews to the blockchain:
First, the smart contract can specify a white-list of addresses allowed to interact with it.
Therefore, for each transaction targeting the contract its sender address (\texttt{msg.sender}) is compared with the list.
Since all addresses (i.e., app users) are unknown beforehand, the contract must implement a method to register additional addresses.
As a drawback, this method can be used to register addresses that did not download the app.
Second, an ERC-20 token\footnote{https://www.ethereum.org/token} can be issued and distributed to app users.
The smart contract methods must require the sender address to own a token before execution.
Further, the contract must ensure that these tokens are untradeable between users.
This alternative also requires tokens to be transferred to reviewers by a central authority.
Third, a private key to sign transactions can be bundled with the app.
By solely white-listening this key's address in the smart contract, only app users are authorized to submit reviews.
To use this key, its passphrase must be included in the app as well.
This introduces a security risk, since key and passphrase can be extracted so that fake reviews can be created from outside of the app.
Additionally, since all users use the same key, we are unable to distinguish them. 
As a consequence, we cannot ensure that each user only submits one review per product version.


\subsection{Reducing Costs of Data Storage}

Another challenge is the storage of review data and its associated costs.
The Ethereum yellow paper \cite{wood2014ethereum} states that a fee of 20,000 gas is required to store a 256-bit word on the blockchain, i.e., 625 gas per byte.
Gas is a unit to measure the computational effort required to perform an action within the Ethereum network.
The cost of a transaction, e.g., storing data on the blockchain, is calculate by multiplying the amount of gas with the gas price.
The gas price is measured in Gwei, where 1 ETH equals 1 billion Gwei.
The price is decided by the miners, who perform the transactions, based on the network conditions.
If transactions specify a too low gas price, these will not be processed by miners.
Vice versa, the higher the price, the faster transactions will be processed.
At the time of writing, in February 2018, the gas price to process a transaction below 5 minutes is 5 Gwei, while the median price of the last 1,500 blocks is 22 Gwei.
The value of one ETH is \$885.

A single 7-day Spotify release for iOS, for example, contains 3,025 reviews with an overall size of 270,110 bytes.
To store this amount of data, an amount of 168,818,750 gas is required.
Considering a gas price of 5 Gwei the storage costs are 0.844 ETH or \$747.
This equals \$0.247 per review.
Using 22 Gwei, the costs are \$3,287 or \$1.09 per review.
The costs can be reduced by defining a low gas price, which consequently requires more time for reviews to be processed.

\begin{table*}[t!]
\centering
\small
\caption{Optimized Trade-Offs and their resulting Configurations of Design Alternatives (A1 refers to Alternative 1).}
\label{tab:tradeoffs}
\begin{tabular}{r|r|r|lllll|l}
\toprule
\multicolumn{3}{c}{\textbf{Trade-Offs}} & \multicolumn{5}{|c}{\textbf{Design Alternatives}} & \multicolumn{1}{|l}{\textbf{Optimi-}}                                                                                                                                                                                \\
\multicolumn{1}{r|}{Security} & \multicolumn{1}{r|}{Trust} & \multicolumn{1}{r|}{Costs} & \multicolumn{1}{l|}{Section 4.1: Submission}       & \multicolumn{1}{l|}{4.2: Authorization}   & \multicolumn{1}{l|}{4.3: Data Storage}     & \multicolumn{1}{l|}{4.4: Transact. Fees}      & 4.5: Retrieval & \multicolumn{1}{l}{\textbf{zed for}}  \\ \midrule
\textbf{Good}                            & Medium                   & Medium & \multicolumn{1}{l|}{A2: Via App (Directly)} & \multicolumn{1}{l|}{A2: ERC-20 Token}      & \multicolumn{1}{l|}{A3: Decentralized} & \multicolumn{1}{l|}{A3: Smart Contract} & A2: Local & Security                         \\ \rowcolor{lightgray}
Poor                             & \textbf{Good}                          & Medium & \multicolumn{1}{l|}{A2: Via App (Directly)} & \multicolumn{1}{l|}{A3: Pool Key (App)} & \multicolumn{1}{l|}{A3: Decentralized} & \multicolumn{1}{l|}{A3: Smart Contract} & A2: Local & Trust                         \\
Good                             & Poor                          & \textbf{Good} & \multicolumn{1}{l|}{A2: Via App (Directly)} & \multicolumn{1}{l|}{A2: ERC-20 Token}      & \multicolumn{1}{l|}{A2: Centralized}   & \multicolumn{1}{l|}{A3: Smart Contract} & A2: Local & Costs                              \\ \rowcolor{lightgray} \bottomrule
\end{tabular}
\end{table*}

\begin{figure}[b]
\includegraphics[width=\columnwidth]{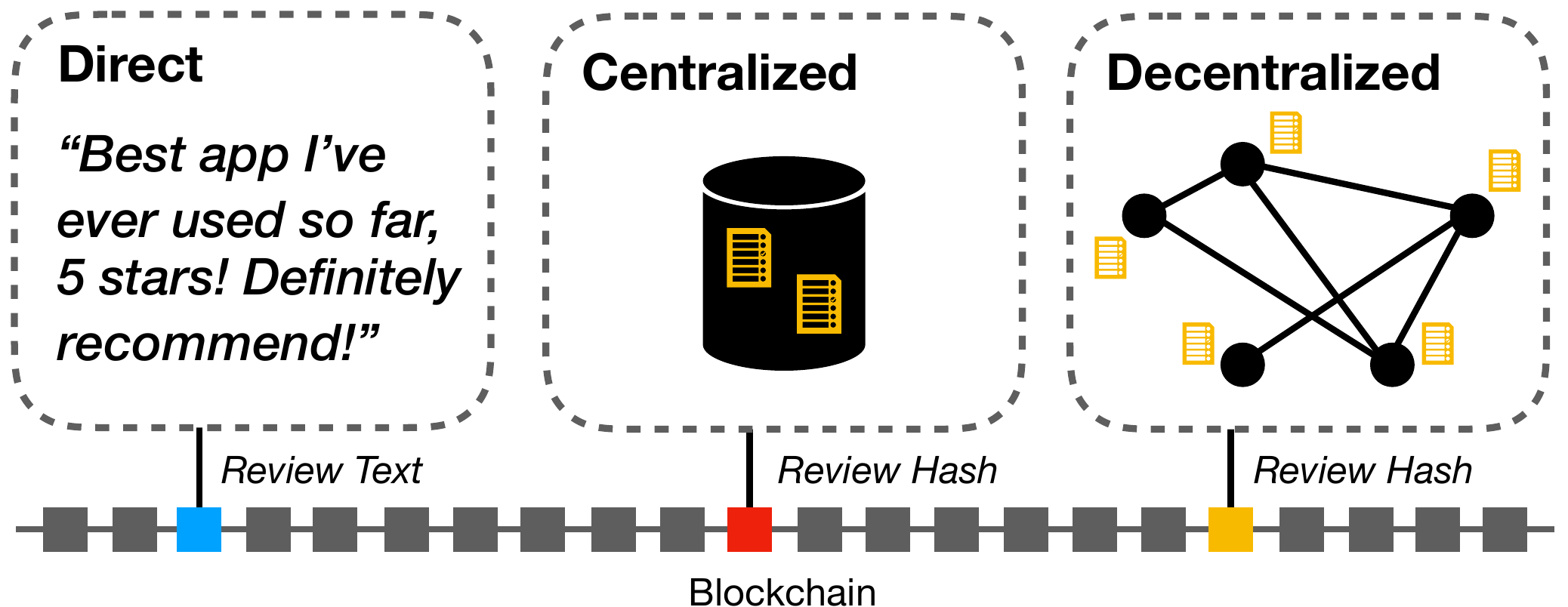}
\caption{Overview of Data Storage.}
\label{fig:datastorage}
\end{figure}

In the following, we discuss three solutions to reduce storage costs:
First, compression algorithms can be used.
Alakuijala et al. \cite{alakuijala2015comparison} introduce Brotli, a compression algorithm including a static dictionary and thereby appropriate for short texts.
Second, instead of the complete review only its hash can be persisted on the blockchain.
The hash length is independent of the review.
This reduces costs of longer reviews.
The actual review can be stored on a central server. 
When accessing reviews, their hashes are compared to those on the blockchain.
However, this design introduces a single point of failure.
Third, a better alternative is to use a a peer-to-peer distributed file system, such as Swarm\footnote{https://github.com/ethereum/go-ethereum/tree/swarm} or InterPlanetary File System (IPFS)\footnote{https://ipfs.io}, see Figure \ref{fig:datastorage}.
These seek to connect all computing devices with the same system of files.
IPFS, for example, provides an immutable, content-addressed block storage model.
Each item can be identified by a hash and accessed using an URL.
By using a distributed file system, the single authority and point of failure is removed.


\subsection{Third-Party Transaction Fee Payments}

When performing a transaction, the sender must pay its fee, i.e., in our case the reviewer.
As our approch requires that submissions are free of costs for reviewers, we present and discuss three design alternatives in the following:

First, as we decided against using a centralized backend to store users' private keys, we do not know their addresses.
To collect these, a website can be deployed for users to (automatically) publish their addresses. 
Afterwards, the addresses can be funded with ETH.
However, this website can be used without submitting a review afterwards.
Second, free-of-charge transactions with a gas price set to zero can be used.
These transactions are very unlikely to be picked by miners.
Therefore, a central authority has to mine all transactions going to the contract's address.
As a drawback, the authority could decide against processing single transactions to, e.g., exclude users writing negative reviews.
Also, the time to process a review highly depends on the mining capacity of the central authority.
Third, to avoid a central authority a smart contract can be used to reward miners.
A miner that processed a transaction pointing to the address of the contract has to prove that the gas price was zero.
The smart contract can then refund the miner for processing the transaction.
The reward is paid from a pool address into which product vendors deposit tokens.
It should equal the median transaction fees, normally paid by transaction senders, so that these are as likely to be picked by miners as regular transactions.


\subsection{Retrieval of Untampered Reviews}

Our aim is to enable users to access reviews on the blockchain in a usable and untampered manner.
Therefore, we deploy a decentralized app (dApp).
This app must fulfill several requirements: 
It must be completely open source,
operate autonomously, and
data must be decentrally stored to avoid central points of failure.
Further, dApps must use tokens for access and to rewards miners, e.g., Gas.

From an implementational perspective, the dApp consists of a frontend and backend.
The frontend can be developed in any language, such as JavaScript.
To fulfill the above mentioned requirements, it is deployed on a decentralized storage such as Swarm or IFPS.
The backend is implemented as a smart contract and deployed on the Ethereum blockchain.
The dApp connects to an Ethereum node using an remote procedure call (RPC) connection.
As the reviews are timely distributed, the node has to have a full replica of the blockchain.
In case a remote node is used, the user has to trust that it returns unmodified results.
To ensure that the dApp reads untampered data from the blockchain, the user has to start a node by herself.
As a drawback, a large amount of data has to be downloaded, which requires space and time, and is unappropriate for mobile devices.


\section{Discussion}

\Cref{tab:tradeoffs} list three possible configurations of design alternatives (one per row).
Each configuration combines different design alternatives with regard to their trade-offs.
The first alternative is optimized for high security.
It submits transactions directly from within the app, uses an ERC-20 token to authorize consumers to submit reviews, and stores data decentrally (e.g., using IFPS or Swarm). 
Further, it uses smart contracts to reward miners processing the review transactions. For retrieval of reviews it uses a decentralized app accessing a local Ethereum node.
For the second alternative trustworthiness is optimized. 
In comparison to the first, it bundles the app with a private pool key to sign transactions, and thereby completely avoids central authorities.
As a drawback, this design reduces the implementation's security as the private key and its passphrase can be extracted from the app.
The last alternative focuses on optimizing costs.
Except for using a centralized database to store the complete reviews (not hashes), it equals the first approach.
Thereby it reduces storage costs, with the drawback of reducing the trustworthiness of reviews, as the database contents can be modified by its operator.

We conclude by highlighting, that depending on the trade-offs to be achieved, such as optimizing for security, trustworthiness, or costs, different configurations of design alternatives need to be chosen.
This can either be a completely decentralized approach, based on blockchain technologies, or a combination of decentralized and centralized design alternatives, such as a blockchain to store reviews hashes and a conventional database to store its contents.

\bibliographystyle{ACM-Reference-Format}
\bibliography{bibliography}

\end{document}